\documentclass[pra,aps,twocolumn,showpacs]{revtex4}
\usepackage{amsmath,graphicx}

\begin{document}

\title{Detection of radio frequency magnetic fields using
nonlinear magneto-optical rotation}

\author{M.\ P.\ Ledbetter}\email{ledbetter@berkeley.edu}
\author{V.\ M.\ Acosta}
\author{S.\ M.\ Rochester}
\author{D.\ Budker}\email{budker@berkeley.edu}
\affiliation{Department of Physics, University of California at
Berkeley, Berkeley, California 94720-7300}
\author{S.\ Pustelny}
\affiliation{Centrum Magnetooptycznych, Instytut Fizyki im. M.
Smoluchowskiego, Uniwersytet Jagiello\'{n}ski, Reymonta 4, 30-059
Krakow, Poland}
\author{V.\ V.\ Yashchuk}
\affiliation{Advanced Light Source Division, Lawrence Berkeley
National Laboratory, Berkeley, California 94720}

\date{\today}

\begin{abstract}
We describe a room-temperature alkali-metal atomic magnetometer for
detection of small, high frequency magnetic fields.  The
magnetometer operates by detecting optical rotation due to the
precession of an aligned ground state in the presence of a small
oscillating magnetic field.  The resonance frequency of the
magnetometer can be adjusted to any desired value by tuning the bias
magnetic field.  We demonstrate a sensitivity of $100\thinspace{\rm
pG/\sqrt{Hz}\thinspace(RMS)}$ in a 3.5 cm diameter, paraffin coated
cell. Based on detection at the photon shot-noise limit, we project
a sensitivity of $20\thinspace{\rm pG/\sqrt{Hz}\thinspace(RMS)}$.

\end{abstract}




\maketitle

\section{Introduction}
Detection of small oscillating magnetic fields is the cornerstone of
experimental techniques such as nuclear magnetic resonance (NMR),
magnetic resonance imaging (MRI), nuclear quadrupole resonance (NQR)
\cite{Garroway01} and has been used in tests of physics beyond the
standard model \cite{Bradley03}. Most atomic magnetometers (for
example, see Refs.\
\cite{Kom03,Budker2000PRA,Acosta06,AlexandrovMag}) are designed to
detect slowly varying magnetic fields and hence are not ideally
suited for the aforementioned applications. In recent work, Savukov
\textit{et al} \cite{Savukov05} demonstrated a tunable,
radio-frequency (RF) alkali vapor magnetometer, achieving a
sensitivity of $20\thinspace{\rm pG/\sqrt{Hz}}$, with the sensor
operating at $190^\circ$C.

Here we present an RF atomic magnetometer based on nonlinear
magneto-optical rotation (NMOR) arising due to the response of an
aligned atomic ground state to a small RF magnetic field near the
Zeeman resonance frequency.  The Zeeman resonance frequency can be
tuned to any desired value by adjusting the bias magnetic field,
yielding sensitivity to signals of arbitrary frequency. The
measurement involves a single low-power light beam ($\sim
50-100\thinspace {\rm \mu W}$), and based on photon shot-noise
limited polarimetry, achieves a sensitivity of about
$20\thinspace{\rm pG/\sqrt{Hz}\thinspace(RMS)}$. The magnetometer
has a bandwidth of about 50-100 Hz depending on light power. Based
on experimentally observed signal-to-noise ratio, we demonstrate a
sensitivity of $100\thinspace{\rm pG/\sqrt{Hz}\thinspace(RMS)}$.
Despite somewhat lower sensitivity than reported in Ref.\
\cite{Savukov05}, for many applications, the magnetometer described
here has the significant advantage that the sensor operates much
closer to room temperature (the highest temperature used in this
work was $48^\circ$C). Additionally, using an aligned state rather
than an oriented state produces smaller external magnetic fields
which can potentially have a back reaction on the sample of
interest. Furthermore, the low power requirements and single beam
arrangement facilitate the use of microfabrication techniques,
promising for the development of compact portable atomic
magnetometers \cite{Schwindt04}.

Such a magnetometer may find application in NQR as suggested in
Ref.\ \cite{Savukov05} where the signal occurs at a fixed resonance
frequency or in NMR spectroscopy where high spectral resolution is
required to observe small splittings of NMR lines, due to, for
example, scalar spin-spin $(J)$ coupling between nuclei of the form
$J\mathbf{I}_1\cdot\mathbf{I}_2$. Such couplings can yield valuable
information on molecular structure \cite{McDermott2002,Appelt2006}
and can be difficult to access in high field environments where the
absolute field homogeneity and differences in diamagnetic
susceptibility limit the spectral resolution. Hence, recent
attention has been given to performing such measurements in a low
field environment using broadband, low transition-temperature
superconducting quantum interference devices (SQUIDs)
\cite{McDermott2002} or inductive detection \cite{Appelt2006}. As
inductive detection becomes less efficient at low frequencies, the
technique described in this letter offers the possibility of
significant gains in signal-to-noise ratio without requiring
cryogenics.

\section{Radio frequency NMOR resonance}
\begin{figure}
    \includegraphics[width=3.3 in]{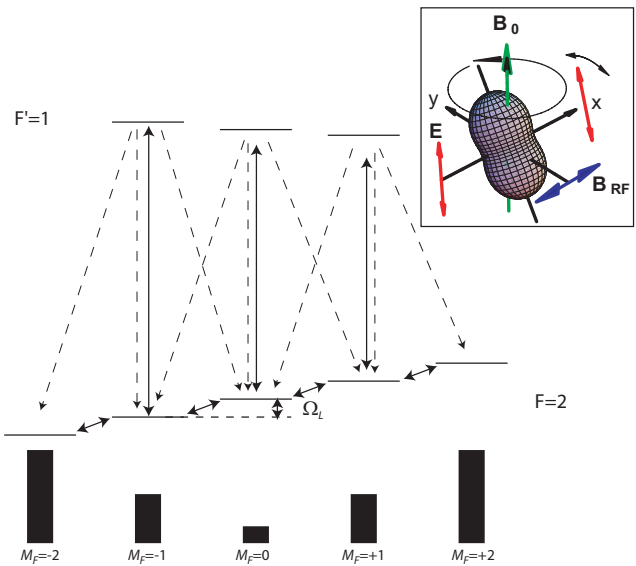}\\
    \caption{ Linearly polarized light, resonant with the D1
    $(F=2\rightarrow F'=1)$ transition, with polarization vector
    along $\mathbf{B}_0$, produces an aligned ground state via
    optical pumping. Double headed vertical arrows indicate laser
    induced transitions between ground and excited states; dashed
    lines indicate transitions due to spontaneous decay. Ground
    state populations, indicated by the solid black bars, are
    schematic only. A small RF magnetic field oscillating close to
    $\Omega_L$, transverse to $\mathbf{B}_0$, establishes coherences
    between neighboring $M_F$ states. Inset: surface whose radius
    represents the probability for finding maximal projection of
    ground state angular momentum along a given direction (see, for
    example Ref.\ \cite{Roc01}).}\label{Fig:Levelscheme}
\end{figure}

An alkali-metal vapor contained in a glass cell with anti-relaxation
coated walls is placed in a $z$ directed bias field $\mathbf{B}_0 =
B_0\mathbf{\hat{z}}$, corresponding to Larmor frequency $\Omega_L=g
\mu_B B_0$ where $\mu_B$ is the Bohr magneton and $g\approx
2/(2I+1)$ is the Land\'e factor. Linearly polarized light
propagating in the $x$ direction with polarization in the $z$
direction, tuned to the D1 ($F=2\rightarrow F'=1$) transition,
passes through the cell, optically pumping an aligned state, as
illustrated in Fig.\ \ref{Fig:Levelscheme}. We apply a small
oscillating magnetic field transverse to the bias field, $B_x = B_1
\cos\omega_{rf} t$ and we work in the regime where $g\mu_B B_1 \ll
\gamma_{rel}$ so that the RF magnetic field induces only ground
state transitions with $|\Delta M_F|=1$.

We begin by assuming the light power is weak enough so that the
saturation parameter relating the optical excitation rate to the
ground state relaxation rate \cite{NMOEreview}
\begin{equation}
    \kappa=\frac{d^2E^2}{\hbar^2
    \Gamma_D\gamma_{rel}}\frac{V_{beam}}{V_{cell}}\label{Eq:sat_param}
\end{equation}
is small compared to unity.  In Eq.\ \eqref{Eq:sat_param}, $d$ is
the electric dipole matrix element of the optical transition, $E$ is
the light electric field, $\Gamma_D$ is the Doppler broadened width
of the optical transition, $\gamma_{rel}$ is the ground state
relaxation rate, $V_{cell}$ is the volume of the cell and $V_{beam}$
is the volume contained within the intersection of beam and cell.
When $\kappa\ll 1$ only the rank 2 (quadrupole) polarization moment
is pumped by linearly polarized light.

The oscillating RF magnetic field can be resolved into components
co- and counter-rotating (parallel or antiparallel to the direction
of Larmor precession respectively), each of magnitude $B_1/2$.
Transforming to the co-rotating frame the counter-rotating component
rapidly averages to zero and the magnetic field is
\begin{equation}
    \mathbf{B}'=\frac{\Omega_L-\omega_{rf}}{g\mu_B}\mathbf{\hat{z}}+
    \frac{B_1}{2}\mathbf{\hat{x}}\thinspace.\label{Eq:rotframefield}
\end{equation}
In steady state, an equilibrium is reached between optical pumping
of alignment along the $z$ axis, precession around $\mathbf{B}'$ and
relaxation, resulting in an aligned state tilted away from the $z$
axis, as shown inset in Fig.\ \ref{Fig:Levelscheme}. When
$\omega_{rf}=\Omega_L$, the $z$ component in Eq.\
\eqref{Eq:rotframefield} vanishes resulting in the maximum angle
between the aligned state and the $z$ axis. When we transform back
into the lab frame, the tilted alignment precesses about the $z$
axis.  The tilted alignment generates optical rotation through
linear dichroism (see, for example, Ref.\ \cite{Kanorsky93}),
maximal when the alignment is in the $yz$ plane and none when it is
in the $xz$ plane, resulting in polarization rotation of the light
beam that is modulated at $\omega_{rf}$. The amplitude of the
polarization rotation is linear in $B_1$ in the range $g\mu_B B_1
\ll \gamma_{rel}$ considered here, as we have verified
experimentally.

The description becomes slightly more complicated for conditions of
high light power and light frequency detuned from optical resonance.
Under these conditions, AC Stark shifts can lead to differential
shifts of the ground state energy levels. In conjunction with
precession in the RF magnetic field, this results in
alignment-to-orientation conversion (AOC) (see Ref.\
\cite{Budker2000PRL} and references therein) in the rotating frame
and a splitting of the RF NMOR resonance as discussed briefly below.
Doppler broadening can also lead to AOC effects, even for resonant
light \cite{Budker2000PRL}. An additional high light power effect is
the generation of hexadecapole (rank 4) polarization moments
\cite{Yashchuk2003}. We find experimentally that optimal sensitivity
is achieved when the saturation parameter is close to unity, and
density matrix calculations indicate that the hexadecapole
contribution to the ground state polarization is small compared to
that of the quadrupole contribution for these conditions.

\section{Experimental setup}
\begin{figure}
  \includegraphics[width=3.4in]{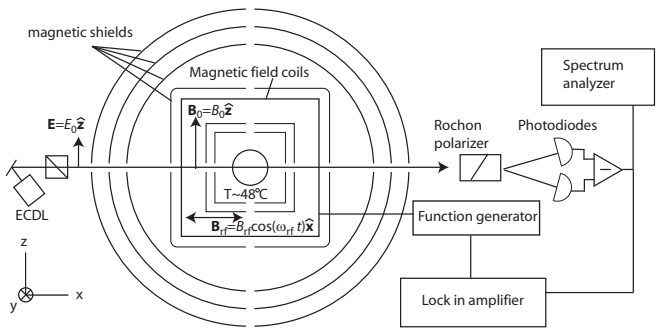}\\
  \caption{Schematic of the experimental setup. An evacuated,
  paraffin coated cell is placed inside a double wall oven.
  Temperature is controlled by flowing warm air through the space
  between the oven walls.  A set of mu-metal layers provides a
  shielded environment and a set of coils maintains a stable,
  homogeneous magnetic field in the $z$ direction. An additional
  coil generates a small oscillating magnetic field in the $x$
  direction. Linearly polarized light from an external cavity diode
  laser passes through the cell and a balanced polarimeter monitors
  the polarization of the light as it exits the cell.
  }\label{Fig:Setup}
\end{figure}

A schematic of the experimental setup is shown in Fig.\
\ref{Fig:Setup}. The measurements reported in this work were
performed with an evacuated, paraffin-coated spherical cell
($3.5\thinspace{\rm cm}$ diameter) containing isotopically enriched
$^{87}$Rb (nuclear spin $I=3/2$). The paraffin coating enables
atomic ground-state polarization to survive tens of thousand wall
collisions \cite{Rob58,Bou66}, leading to ground-state polarization
lifetimes $\tau=1/\gamma_{rel} \approx 160\thinspace{\rm ms}$ in a
10~cm diameter cell \cite{Bud98}. The cell is placed inside a
double-wall oven, temperature-controlled by flowing warm air through
the space between the walls of the oven so that the optical path is
unperturbed.   A set of four nested $\mu$-metal layers provides a
magnetically shielded environment, with a shielding factor of
approximately $10^6$ \cite{shieldref}. Inside the innermost shield
(cubic in profile) is a set of coils for the control of all three
components of the magnetic field. Image currents in the magnetic
shields create an ``infinitely'' long solenoid. The atoms traverse
the cell many times during the course of one relaxation period,
effectively averaging the magnetic field over the cell, leaving our
measurements insensitive to field gradients \cite{Pustelnygrad}. We
apply a static magnetic field $B_0$ in the $z$ direction and a small
oscillating magnetic field $B_1 \cos(\omega_{rf}t)$ in the $x$
direction (unless stated otherwise, $B_1= 110\thinspace {\rm nG}$
and $B_0\approx 10\thinspace{\rm mG}$). A well collimated beam with
diameter $\approx 3\thinspace{\rm mm}$ from an external-cavity diode
laser, propagates in the $x$ direction with polarization vector in
the $z$ direction. Unless stated otherwise, these measurements were
performed with the light tuned to the center of the $F=2\rightarrow
F'=1$ transition (henceforth referred to as optical resonance). The
polarization of the light leaving the cell is monitored using a
balanced polarimeter and detected synchronously using a lock-in
amplifier. Number density was determined by monitoring the
transmission of a low-power beam through the cell as a function of
laser frequency. For the measurements reported here, the cell
temperature was $48^\circ$C and the measured number density was
$n=7\times 10^{10}$ (within 20\% of that expected from the saturated
vapor pressure at this temperature), corresponding to approximately
1 absorption length for resonant light.

\section{Experimental results and discussion}

\begin{figure}
    \includegraphics[width=3.4in]{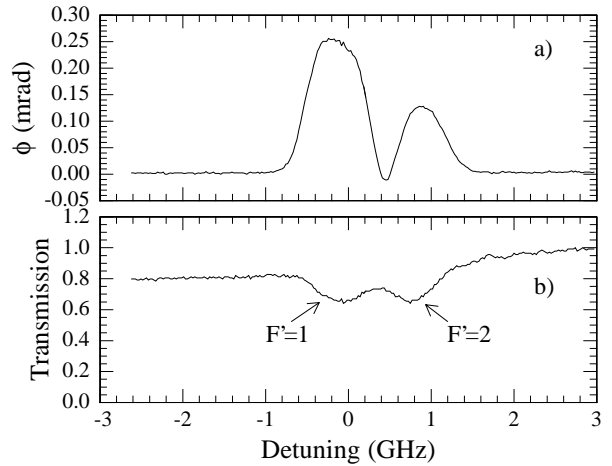}\\
    \caption{ a: Synchronously detected optical rotation and b:
    transmission spectra as a function of light frequency for a light
    power of $60\thinspace{\rm \mu W}$ and $\omega_{rf}=\Omega_L$.
    }\label{Fig:rotation_vs_freq}
\end{figure}

In Fig.\ \ref{Fig:rotation_vs_freq}a we plot the in-phase component
of the synchronously detected optical rotation as a function of
light frequency for $\omega_{rf}=\Omega_L$. For these data, the
light power was $60\thinspace{\rm \mu W} (850 \thinspace{\rm \mu
W/cm^{2}})$. In Fig.\ \ref{Fig:rotation_vs_freq}b we plot the
partially saturated transmission curve under the same experimental
conditions. The background slope of the transmission curve is due to
varying laser intensity as the diode laser feedback grating is
swept.  The largest optical rotation occurs for light tuned near the
center of the $F=2\rightarrow F'=1$ transition, similar to
observations of non-linear Faraday rotation induced by a static
magnetic field \cite{Bud02}. At the light powers for which we
obtained optimal sensitivity on the $F=2$ component, optical
rotation on the $F=1$ component was at least an order of magnitude
smaller than that produced by the $F=2$ component.

\begin{figure}
  \includegraphics[width=3.4in]{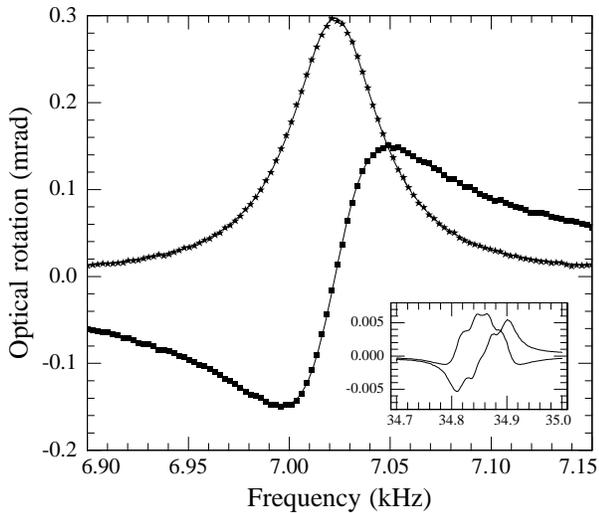}\\
  \caption{In-phase (stars) and quadrature (squares), synchronously
  detected optical rotation signal as a function of the frequency of
  the applied oscillating magnetic field.  Overlaying the data
  (solid lines) are fits to a single dispersive or absorptive
  Lorentzian. For these data, the light was tuned approximately to
  the center of the $F=2\rightarrow F'=1$ transition and the power
  was $40\thinspace{\rm \mu W}$. Inset: Optical rotation signals for
  light power of $360\thinspace{\rm \mu W}$, cell temperature of
  $20^\circ$C and optical detuning far to the red of the
  $F=2\rightarrow F'=1$ transition. AC Stark shifts result in a
  splitting of the resonance. Units are the same as in the main
  panel. The overall shift of the resonance is different from that
  in the main panel because the bias field differed by a factor of
  5.}\label{Fig:RF_resonance}
\end{figure}
In the main panel of Fig.\ \ref{Fig:RF_resonance} we show the
synchronously detected in-phase (stars) and quadrature (squares)
components of optical rotation for light tuned to optical resonance
and incident light power of $40\thinspace{\rm \mu W}$. Overlaying
the in-phase (quadrature) component is a fit to a single absorptive
(dispersive) Lorentzian. As mentioned previously, under conditions
of high light power and detuning far from optical resonance,
differential AC Stark shifts can lead to a modification of the
quantum beat frequency for different $\Delta M_F = 1$ transitions
resulting in a splitting of the resonance, as shown inset in Fig.\
\ref{Fig:RF_resonance}.  The overall shift of the resonance compared
to that shown in the main panel is because the bias magnetic field
differed by a factor of 5. AC Stark shifts were discussed in some
detail in Ref.\ \cite{Acosta06} in the context of NMOR with
frequency modulated light. It was found that magnetometric
sensitivity was reduced when the resonance was split and hence we do
not focus on this behavior any further. However, we point out that
AC Stark shifts may provide some degree of optical tunability of the
Zeeman resonance and we will explore this possibility in future
work.

\begin{figure}
  \includegraphics[width=3.4in]{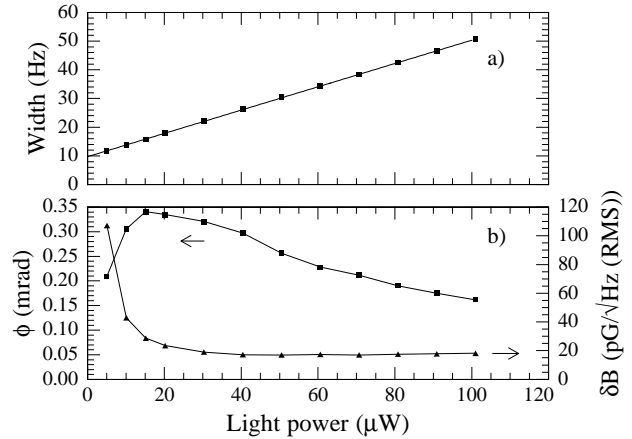}\\
  \caption{a: Half-width at half-maximum of the in-phase component
  of the RF NMOR resonance shown in Fig.\ \ref{Fig:RF_resonance} as a
  function of light power.  The solid line overlaying the data is a
  linear fit. b: The amplitude (squares) of the RF Zeeman resonance
  as a function of light power against the left axis.  Against the
  right axis we plot the projected sensitivity (triangles) based on
  the amplitude of the resonance and photon shot-noise limited
  polarimetry. Points are joined as a guide for the
  eye.}\label{Fig:width_amp}
\end{figure}

In Fig.\ \ref{Fig:width_amp}a we plot $\Delta\nu$, the half-width at
half-maximum, of the in-phase component of the RF NMOR resonance, as
a function of light power (the distance from the center of the
resonance to the extrema of the quadrature signal is also given by
$\delta \nu$). Overlaying the data is a linear fit with zero-power
width $\Delta\nu_0=9.7\thinspace{\rm Hz}$. The intrinsic
polarization relaxation rate $\gamma_{rel}$ is related to
$\Delta\nu_0$ via $\gamma_{rel}=2 \pi \Delta \nu_0$
\cite{nmor_RF_width_comp}. Ground state relaxation in paraffin
coated cells is typically dominated by electron randomization during
collisions with the cell walls and through alkali-alkali spin
exchange collisions (see for example \cite{Budker2005,Graf05} and
references therein). The relaxation rate for the latter process is
given by \cite{spinexref}
\begin{equation}\label{Eq:SE_relaxation}
    \gamma_{SE}\approx\frac{1}{2}\sigma_{SE}v_{rel}n= 2\pi\cdot 6 \cdot 10^{-11}
    \thinspace{\rm cm^{3}\thinspace Hz}\cdot n.
\end{equation}
Here $\sigma_{SE}\approx 2\cdot 10^{-14}\thinspace{\rm cm^{2}}$ is
the spin-exchange cross section, and $v_{rel}=\sqrt{8kT/\pi\mu}$ is
the average relative speed of the atoms, $\mu$ is the reduced mass.
The factor of $1/2$ in Eq.\ \eqref{Eq:SE_relaxation} represents the
approximate nuclear ``slowing down'' factor appropriate for a spin
3/2 nucleus.  For a density $n=7\cdot 10^{10}\thinspace{\rm
cm^{-3}}$, Eq.\ \eqref{Eq:SE_relaxation} gives
$\gamma_{SE}=2\pi\cdot 4.2\thinspace{\rm Hz}$, roughly a factor of 2
smaller than the experimentally measured relaxation rate.  We
attribute the excess relaxation to collisions with the walls.

In Fig.\ \ref{Fig:width_amp}b we plot the amplitude of the RF NMOR
resonance shown in Fig.\ \ref{Fig:RF_resonance} (defined as the
maximum of the in-phase component) against the left axis as a
function of input light power.  The amplitude increases as a
function of light power for low light power, until reaching a
maximum at around $15\thinspace{\rm \mu W}$ corresponding to $\kappa
\approx 1.5$. Beyond saturation the amplitude decreases due to light
broadening. Against the right axis in Fig.\ \ref{Fig:width_amp}b we
plot the sensitivity of the magnetometer assuming a photon
shot-noise limited polarimeter sensitivity
$\delta\phi_{ph}=1/(2\sqrt{\Phi_{ph}})$ where $\Phi_{ph}$ is the
number of photons per second exiting the cell. Light power is
measured both before and after the beam passes through the cell to
accurately take into account optical absorption. Optimum sensitivity
of about $20\thinspace {\rm pG/\sqrt{Hz}\thinspace(RMS)}$ is reached
at about $40-50\thinspace{\rm \mu W}$ input light power
\cite{bandwidth}. The bandwidth of the magnetometer (the range of
frequencies over which the signal is greater than half the value
when $\omega_{rf}=\Omega_L$) is given by the full-width at
half-maximum, about 50 Hz at $40\thinspace{\rm \mu W}$. By
increasing light power to $100\thinspace{\rm \mu W}$ the bandwidth
may be doubled with little loss in sensitivity.

Atomic shot noise places a fundamental limit on the magnetometer
sensitivity, given by
\begin{equation}\label{Eq:atomshotnoise}
    \delta B_{atom} = 2\frac{1}{g
    \mu_B}\frac{1}{\sqrt{nV_{cell}\tau}}\approx 4
    \thinspace{\rm pG/\sqrt{Hz}\thinspace(RMS)}\thinspace.
\end{equation}
The factor of 2 is due to the fact that the magnetometer is
sensitive only to the co-rotating component of an oscillating
magnetic field. For an optimized magnetometer, one would expect that
photon shot noise (adding in quadrature to the atom shot noise)
would be comparable to the atomic shot noise \cite{Auzinsh} and
hence Eq.\ \eqref{Eq:atomshotnoise} must be multiplied by a factor
of $\sqrt{2}$ for a fair comparison. Thus, for optimized light
power, the sensitivity based on photon shot noise in Fig.\
\ref{Fig:width_amp} is within a factor of 2 or 3 of the fundamental
limit. This indicates that the optical pumping and probing scheme in
this work is an efficient method for detecting spin precession,
despite the fact that atoms can be pumped into the optically dark
$F=1$ state. Some improvement in sensitivity may be achieved by
increasing number density, however, in the regime where
spin-exchange is the dominant relaxation process, $\tau$ scales
inversely with density so that the shot-noise limit is independent
of density at sufficiently high densities. We point out that
increasing the number density can yield higher bandwidths.

A complete theoretical treatment of NMOR is difficult because of the
presence of hyperfine structure, Doppler broadening, velocity mixing
and evolution in the dark. Following the method outlined in Ref.\
\cite{NMOEreview} we performed a simplified steady-state density
matrix calculation on an $F=2\rightarrow F'=1$ transition which
neglects these issues, but qualitatively reproduces the salient
features of our experimental data. The hamiltonian is written in the
rotating-wave approximation, neglecting terms counter rotating at
either the optical or radio frequencies. The density matrix
evolution equations are then formed, including terms describing
spontaneous decay of the excited state, and atoms entering and
leaving the interaction region (transit relaxation), and solved
numerically. For conditions of high light power and detuning far
from optical resonance, the model reproduces the splitting of the RF
NMOR resonance shown inset in Fig.\ \ref{Fig:RF_resonance}. When the
light is tuned to optical resonance, a single feature is observed in
the RF dependence of the optical rotation. The calculated power
dependence of the amplitude of the resonance is similar to the
experimentally observed behavior shown in Fig.\ \ref{Fig:width_amp}.
The model indicates that for light power that maximizes optical
rotation (saturation), the hexadecapole contribution to the ground
state polarization is small, (roughly 10\%) compared to that of the
quadrupole contribution.

\section{RF magnetometer performance}
\begin{figure}
  \includegraphics[width=3.4 in]{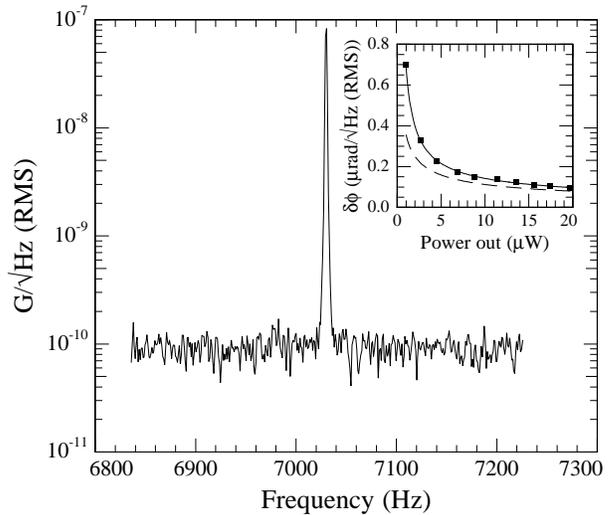}\\
  \caption{RF magnetic noise spectrum.  The large peak is an applied
  field of $83\thinspace{\rm nG\thinspace(RMS)}$ at $\Omega_L$.
  Light power was $40\thinspace{\rm \mu W}$. Shown inset is the
  polarimeter noise (squares) as a function of light power incident
  on the polarimeter. The solid line is a fit base on Eq.\
  \eqref{Eq:polarimeter_noise} and the dashed line represents photon
  shot noise.}\label{Fig:SNR_Sensitivity}
\end{figure}
In Fig.\ \ref{Fig:SNR_Sensitivity} we plot the noise spectrum as
measured by an SRS770 spectrum analyzer at the output of the
balanced polarimeter. The large peak is an applied field of
$83\thinspace {\rm nG\thinspace(RMS)}$ to calibrate the
magnetometer. Baseline noise is about $100\thinspace {\rm
pG/\sqrt{Hz}\thinspace (RMS)}$, falling somewhat short of the
sensitivity estimates based on photon shot noise detection in Fig.\
\ref{Fig:width_amp}b. Deformation of the beam by the cell resulted
in a factor of roughly 5 loss of light power, pointing towards an
obvious improvement for future work. In order to asses the
performance of the polarimeter, shown inset in Fig.\
\ref{Fig:SNR_Sensitivity} is the measured noise floor (squares) as a
function of light power incident on the polarimeter. The dashed line
represents photon shot-noise
$\delta\phi_{ph}=1/(2\sqrt{\Phi_{ph}})=0.35\cdot \thinspace{\rm\mu
rad\sqrt{\mu W}/\sqrt{Hz}\thinspace (RMS)}$. Polarimeter noise can
be parameterized by
\begin{equation}\label{Eq:polarimeter_noise}
    \delta\phi=\sqrt{\zeta_{ph}^2/P+\zeta_{amp}^2/P^2}.
\end{equation}
Here $P$ is the power incident on the polarimeter and $\zeta_{ph}$
and $\zeta_{amp}$ parameterize photon shot noise and amplifier noise
respectively. The solid line overlaying the data is a fit based on
Eq.\ \eqref{Eq:polarimeter_noise}, resulting in
$\zeta_{amp}=0.55\thinspace{\rm \mu rad\thinspace \mu
W/\sqrt{Hz}\thinspace (RMS)}$ and $\zeta_{ph}=0.41 \thinspace{\rm
\mu rad\sqrt{\mu W/Hz}\thinspace (RMS)}$, close to the theoretically
predicted value. Hence, amplifier noise is the dominant contribution
for incident light power less than about $2\thinspace{\rm \mu W}$
and photon shot-noise dominates for higher light power.

\section{Conclusion}
In conclusion, we have demonstrated a simple atomic magnetometric
technique for the measurement of small RF magnetic fields based on a
ground state Zeeman resonance and detection of non-linear magneto
optical rotation in an alkali-metal vapor. Based on photon shot
noise detection we estimate a sensitivity of approximately
$20\thinspace{\rm pG/\sqrt{Hz}\thinspace (RMS)}$ in a 3.5 cm
diameter cell with a bandwidth of approximately 100~Hz. Estimates of
the atom shot noise limit are within a factor of 2-3 of this limit,
confirming that the present optical pumping and probing scheme is an
efficient method for probing spin precession. Based on the actual
signal to noise ratio, we have achieved a sensitivity of about
$100\thinspace{\rm pG/\sqrt{Hz}\thinspace (RMS)}$. With several
technical improvements, we anticipate a factor of 3-5 gain in
sensitivity. Optimization of number density may yield some further
gains in sensitivity as well as bandwidth. The magnetometer operates
near room temperature, making it particularly attractive for
applications in NMR.  One possible such application is the
measurement of a scalar, electron-mediated nuclear spin-spin
coupling which can yield valuable information on molecular
structure.  The authors thank J. Higbie for useful comments and
discussions.  This work is supported by an ONR MURI program and KBN
grant \# 1 P03B 102 30.


\begin{thebibliography}{99}
\bibitem{nmor_RF_width_comp} For NMOR in the Faraday geometry (see, for example, Ref.\
\cite{Bud98}), the relaxation rate is $\gamma_{rel} = g \mu_B \Delta
B $ where $\Delta B$ represents \textit{twice} the distance from the
center of the resonance to the extrema of the optical rotation. In
the present work, $\Delta\nu$ represents the distance from the
center of the resonance to the extrema of the optical rotation. The
reason for this difference is that the angle between the aligned
state and the initial light polarization vector is always small in
this work, in contrast to NMOR in the Faraday geometry.
\bibitem{bandwidth}  Estimates are based on a bandwidth of 1 Hz
corresponding to a measurement time of 0.5 s.
\end{thebibliography}
\end{document}